\documentstyle[pra,aps]{revtex}
\draft

\begin{document}

\title{Two-Level Atom in an Optical Parametric Oscillator: Spectra of 
Transmitted and Fluorescent Fields in the Weak Driving Field Limit}
\author{J. P. Clemens$^{*}$, P. R. Rice$^{*}$, P. K. Rungta$^{*}$, 
and R. J. Brecha$^{\#}$}
\address{$^{\#}$Dept. of Physics, Univ. of Dayton, Dayton, OH 45469\\
$^{*}$Dept. of Physics, Miami Univ., Oxford, OH 45056}

\date{\today}
\maketitle

\begin{abstract}
We consider the interaction of a two-level atom inside an optical parametric 
oscillator. In the weak driving field limit, we essentially have an 
atom-cavity system driven by the occasional pair of correlated photons, or 
weakly squeezed light. We find that we may have holes, or dips, in the 
spectrum of the fluorescent and transmitted light. This occurs even in the 
strong-coupling limit when we find holes in the vacuum-Rabi doublet. Also, 
spectra with a sub-natural linewidth may occur. These effects disappear for 
larger driving fields, unlike the spectral narrowing obtained in resonance 
fluorescence in a squeezed vacuum; here it is important that the squeezing 
parameter $N$ tends to zero so that the system interacts with only one 
correlated pair of photons at a time. We show that a previous explanation for 
spectral narrowing and spectral holes for incoherent scattering is not 
applicable in the present case, and propose a new explanation.  We attribute 
these anomalous effects to quantum interference in the two-photon scattering 
of the system.
\end{abstract}

\section{Introduction}
Much effort in quantum optics has focused on the generation of nonclassical 
light, predominantly antibunched and quadrature squeezed light. As this 
became realizable in the lab, it became important to consider the interaction 
of optical systems with nonclassical light, i.e. shine nonclassical light on 
optical systems and see how they respond. In the case of quadrature squeezed 
light driving a system one finds spectral narrowing in certain features, and 
usually the narrowing is proportional to the size of the squeezing parameter 
$N$\cite{Gardiner,HJC,HJC2}. In practice it is difficult to ensure that all 
modes the atom(s) couple to are squeezed, so a cavity is sometimes used in 
practice. Here, for a small microcavity, the atom may couple much more 
strongly to the cavity field mode than the vacuum modes, and driving the 
cavity with squeezed light can then induce the desired narrowing. The limit 
here is that the cavity enhancement of the linewidth can be squeezed away, but 
the resulting linewidth is subnatural only if the decay rate to all vacuum 
modes is appreciably less than the free space decay rate \cite{PL}. This is 
difficult to achieve in practice, unless the cavity subtends a substantial 
fraction of $4\pi SR$. It was then suggested by Jin and Xiao \cite{Jin,Jin2} 
that the atom could be placed inside the source of the squeezing. They 
considered phase/intensity bistability in the case of a two-level atom inside 
an optical parametric oscillator. Further they considered the spectrum of 
squeezing and incoherent spectra for that system. It was decided that it would 
be fruitful to examine this system in the weak driving field limit.

In section 2 we examine the physical system under consideration. The 
transmitted spectrum is calculated and discussed in section 3. The spectrum of 
the fluorescent light is considered in section 4. In section 5 we consider the 
physical explanation of the anomalous spectra we see, and we conclude in 
section 6.

\section{Physical System}
We consider a single two-level atom inside an optical cavity, which also 
contains a material with a $\chi^{(2)}$ nonlinearity. The atom and cavity are 
assumed to be resonant at $\omega$ and the system is driven by light at 
$2\omega$. The system is shown in Figure 1. The interaction of this driving 
field with the nonlinear material produces light at the sub-harmonic $\omega$. 
This light consists of correlated pairs of photons, or quadrature squeezed 
light. In the limit of weak driving fields, these correlated pairs are created 
in the cavity and eventually two photons leave the cavity through the end 
mirror or as fluorescence out the side before the next pair is generated. 
Hence we may view the system as an atom-cavity system driven by the occasional 
pair of correlated photons. In the language of squeezed light, we are 
interested in the limit $N\rightarrow 0$. As $N$ is increased the effects we 
consider here vanish. We wish to understand these effects in terms of photon 
correlation's rather than the usual effects of quadrature squeezed light, 
where typically the largest nonclassical effects are seen in the large $N$ 
limit. The system is described by a master equation in Lindblad form
\begin{equation}
\dot \rho =-i\hbar \left[ {H,\rho } \right]+{\cal L}_{diss}\rho \equiv 
{\cal L}\rho
\end{equation}
where the system Hamiltonian is
\begin{equation}
H=i\hbar F(a^{\dagger ^2}-a^2)+i\hbar g\;(a^\dagger \sigma _--a\sigma _+) 
+\hbar \omega (a^\dagger a+{1 \over 2}\sigma _z)
\end{equation}
Here, $g$ is the usual Jaynes-Cummings atom-field coupling in the rotating 
wave and dipole approximations. The cavity-mode volume is $V$, and the atomic 
dipole matrix element connecting ground and excited states is $\mu_d$. The 
effective two-photon driving field $F$  is proportional to the intensity 
$I_{in}(2\omega_0)$ of a driving field at  twice the resonant frequency of the 
atom (and resonant cavity)  and the $\chi^{(2)}$  of the nonlinear crystal in 
the cavity, as
\begin{equation}
F=-i\kappa _{in}\left( {{\cal{F} \over \pi }} \right)\sqrt 
{{{\varepsilon _0V\;T} \over {\hbar \omega }}}e^{i\phi }\chi ^{(2)}I_{in} 
(2\omega )
\end{equation}

The cavity finesse is $\cal{F}$, $T$ and $\phi$ are the intensity transmission 
coefficient and phase change at the input mirror. We also have 
$\kappa_{in}=cT/L$ as the cavity field loss rate through the input mirror. The 
transmission of the input mirror is taken to be vanishingly small, with a 
large $I_{in}(2\omega_0)$  so that $F$ is finite. Hence we effectively 
consider a single ended cavity. The dissipative Liouvillian describing loss 
due to the leaky end mirror and spontaneous emission out the side of the 
cavity is
\begin{eqnarray}
L_{diss}\rho &=&{\gamma  \over 2}(2\sigma _-\rho \sigma _+ -\sigma _
+\sigma _-\rho -\rho \sigma _+\sigma _-)\nonumber \\
&&+\kappa (2a\rho a^\dagger  -a^\dagger a\rho -\rho a^\dagger a)
\end{eqnarray}

Here $\gamma$ is the spontaneous emission rate to all modes other than the 
privileged  cavity mode, hereafter referred to as the vacuum modes. The field 
decay rate of the cavity at the output mirror is $\kappa$. As we are working 
in the
weak driving field limit, we only consider states of the system with up to 2 
quanta, i.e.
\begin{equation}
|00\rangle ,\;|10\rangle ,\;|01\rangle ,\;|20\rangle ,\;|11\rangle
 \label{eq:basisreal}
\end{equation}
Here, the first index corresponds to the number of energy quanta in the
atoms (0 for ground state, and 1 for excited state), and the second 
corresponds to the excitation of the field (n=number
of quanta).  In this basis set we have the following equations for density 
matrix elements

\begin{mathletters}
\begin{eqnarray}
\dot{\rho}_{0,-;0,-}&=&\gamma \rho_{0,+;0,+}+2\kappa \rho_{1,-;1,-}-2 
{\sqrt{2}}F \rho_{0,-;2,-} \\
\dot{\rho}_{0,+;0,+}&=&-\gamma \rho_{0,+;0,+}+2\kappa \rho_{1,+;1,+}-2g 
\rho_{0,+;1,-}\\
\dot{\rho}_{1,-;1,-}&=&\gamma \rho_{1,+;1,+} -2\kappa \rho_{1,-;1,-}+4\kappa 
\rho_{2,-;2,-}+2g \rho_{0,+;1,-}\\
\dot{\rho}_{1,+;1,+}&=&-(\gamma+2\kappa)\rho_{1,+;1,+}-2\sqrt{2} g
\rho_{1,+;2,-}\\
\dot{\rho}_{2,-;2,-}&=&-4\kappa \rho_{2,-;2,-}+2\sqrt{2}F\rho_{0,-;2,-}+2
\sqrt{2}g\rho_{1,+;2,-}\\
\dot{\rho}_{0,-;1,+}&=&-(\gamma/2+\kappa)\rho_{0,-;1,+}-\sqrt{2}g
\rho_{0,-;2,-}-\sqrt{2}F\rho_{1,+;2,-}\\
\dot{\rho}_{0,-;2,-}&=&-2\kappa \rho_{0,-;2,-}+2\sqrt{2}F(\rho_{0,-;0,-}-
\rho_{2,-;2,-})+\sqrt{2}g\rho_{0,-;1,+}\nonumber \\ 
&&+\sqrt{2}F\rho_{1,+;2,-}\label{rhoeq}\\
\dot{\rho}_{0,+;1,-}&=&-(\gamma/2+\kappa \rho_{0,+;1,-} -g(\rho_{1,-;1,-}-
\rho_{0,+;0,+})\nonumber \\
&&+2\sqrt{2}\kappa \rho_{1,+;2,-}\\
\dot{\rho}_{1,+;2,-}&=&-(\gamma/2+3\kappa)\rho_{1,+;2,-}+\sqrt{2}F
\rho_{0,-;1,+}\nonumber \\ &&-\sqrt{2}g(\rho_{2,-;2,-}-\rho_{1,+;1,+})
\end{eqnarray}
\end{mathletters}

The other density matrix elements are not driven by the external field and 
couple only  to themselves, hence if they are initially
zero, they remain zero for all time. In the weak field limit, one might 
expect that the population of the ground state is of order unity. With this 
in mind, examine equation \ref{rhoeq}. Here we see that by taking 
$\rho_{0,-;0,-}\approx 1$ and $\rho_{0,-;0,-} \gg \rho_{2,-;2,-}$, we have
\begin{equation}
\dot{\rho}_{0,-;2,-}=-2\kappa \rho_{0,-;2,-}+2\sqrt{2}F+\sqrt{2}g
\rho_{0,-;1,+}+\sqrt{2}F\rho_{1,+;2,-}
\end{equation}

Here we see that $\rho_{0,-;2,-}$ is driven by a term of order $F$. This leads 
us to propose the $\rho_{0,-;1,+}$ and $\rho_{0,-;2,-}$ scale as $F$ in the 
weak field limit. Carrying this process out in a self-consistent matter we 
arrive at the scalings
\begin{mathletters}
\begin{eqnarray}
\rho_{0,-;0,-}&\approx &1\\
\rho_{0,+;0,+}&\approx &F^2\\
\rho_{1,-;1,-}&\approx &F^2\\
\rho_{1,+;1,+}&\approx &F^2\\
\rho_{2,-;2,-}&\approx &F^2\\
\rho_{0,-;1,+}&\approx &F\\
\rho_{0,-;2,-}&\approx &F\\
\rho_{0,+;1,-}&\approx &F^2\\
\rho_{1,+;2,-}&\approx &F^2
\end{eqnarray}
\end{mathletters}
 These scaling make sense physically, as $\rho_{2,-;2,-}$ is a population 
 driven by $F$ in the Hamiltonian and is then proportional to $F^2$ to first 
 order. The Jaynes-Cummings coupling $g$ then couples $\rho_{2,-;2,-}$ to 
 $\rho_{1,+;1,+}$, so both two-photon
state populations scale as $F^2$. Spontaneous emission and cavity decay are 
then responsible for coupling the two-photon states to the one-photon states, 
making $\rho_{0,+;0,+}$ and $\rho_{1,-;1,-}$ of order $F^2$. The coherences 
$\rho_{0,-;2,-}$ and $\rho_{0,-;1,+}$ are driven directly by $F$. Finally the 
coherence $\rho_{1,+;2,-}$ is driven by the 
population of the two-photon states, and hence is proportional to $F^2$.
Keeping
terms to lowest order in $F$, the relevant equations become
\begin{mathletters}
\begin{eqnarray}
\dot{\rho}_{0,-;0,-}&=&0 \\
\dot{\rho}_{0,+;0,+}&=&-\gamma \rho_{0,+;0,+}+2\kappa \rho_{1,+;1,+}-2g 
\rho_{0,+;1,-}\\
\dot{\rho}_{1,-;1,-}&=&\gamma \rho_{1,+;1,+} -2\kappa \rho_{1,-;1,-}+4\kappa 
\rho_{2,-;2,-}+2g \rho_{0,+;1,-}\\
\dot{\rho}_{1,+;1,+}&=&-(\gamma+2\kappa)\rho_{1,+;1,+}-2\sqrt{2} g
\rho_{1,+;2,-}\\
\dot{\rho}_{2,-;2,-}&=&-4\kappa \rho_{2,-;2,-}+\sqrt{2}F\rho_{0,-;2,-}+2
\sqrt{2}g\rho_{1,+;2,-}\\
\dot{\rho}_{0,-;1,+}&=&-(\gamma/2+\kappa)\rho_{0,-;1,+}-\sqrt{2}g
\rho_{0,-;2,-}-\sqrt{2}F\rho_{1,+;2,-}\\
\dot{\rho}_{0,-;2,-}&=&-2\kappa \rho_{0,-;2,-}+\sqrt{2}F+\sqrt{2}g
\rho_{0,-;1,+}\\
\dot{\rho}_{0,+;1,-}&=&-(\gamma/2+\kappa \rho_{0,+;1,-} 2g(\rho_{1,-;1,-}-
\rho_{0,+;0,+})\nonumber \\
&&+2\sqrt{2}\kappa \rho_{1,+;2,-}\\
\dot{\rho}_{1,+;2,-}&=&-(\gamma/2+3\kappa)\rho_{1,+;2,-}+\sqrt{2}F
\rho_{0,-;1,+}\nonumber \\ &&-\sqrt{2}g(\rho_{2,-;2,-}-\rho_{1,+;1,+})
\end{eqnarray}
\end{mathletters}
In what follows, these equations are numerically solved for the steady-state 
density matrix elements of the system. We note
here that $\langle a\rangle_{ss} = \rho_{0,-;1,-}+\rho_{1,-;2,-}=0$, but 
$\langle a^\dagger a\rangle_{ss} =\rho_{1,-;1,-}+
\rho_{1,+;1,+}+2\rho_{2,-;2,-} \approx F^2$. These results hold in the weak 
field limit, but the mean intracavity field is also zero for 
arbitrary driving field states. 

\section{Optical Spectrum of the Transmitted Light}
We now turn our attention to a calculation of the spectrum of squeezing, and 
incoherent spectrum; we consider both transmitted and 
fluorescent light fields.
For the transmitted spectrum, in a rotating frame such that $\omega=0$ 
corresponds to the simultaneous cavity and atomic resonances,  we have
\begin{eqnarray}
I_{tr}(\omega )&=& \int\limits_{-\infty }^\infty  {d\tau \,e^{i\omega \,
\tau }}\left\langle {a^\dagger(0)\,a\,(\tau )} \right\rangle 
\nonumber \\
&=&2\Re \int\limits_0^\infty  {d\tau \,e^{i\omega \,\tau }}\left\langle 
{a^\dagger(0)\,a\,(\tau )} \right\rangle \nonumber \\
&=&2\pi \langle a\rangle_{ss} \langle a^\dagger \rangle_{ss} \delta (\omega) 
+\nonumber \\
&&2\Re \int\limits_0^\infty  {d\tau \,e^{i\omega \,\tau }}\left\langle {
\Delta a^\dagger(0)\,\Delta a\,(\tau )} \right\rangle
\end{eqnarray}

The first term is due to elastic scattering, and is zero here, as $\langle 
a\rangle_{ss}=0$. The second term is the incoherent, or inelastic spectrum, 
and is due to two photon scattering events. For an optical system driven at 
$\omega$ by a field of strength $E$, the coherent spectrum is usually 
proportional to the driving intensity $E^2$, and the incoherent spectrum is 
proportional to the square of the intensity or $E^4$. Here however, the 
external classical driving field produces pairs of photons via the 
$\chi^{(2)}$ nonlinearity of the intracavity crystal. Thus there are no single 
photon scattering events and no coherent scattering spike, and the
height of the incoherent spectrum depends linearly on $I_{in}(2\omega_0) 
\propto F^2$

By the quantum regression theorem we have
\begin{equation}
\left\langle {a^\dagger (0)\,a\,(\tau )} \right\rangle =tr\left\{ {a(0)
\,A(\tau )} \right\}=\sum\limits_{i,n} {\sqrt {n+1}\,\left\langle {i,n+1\,|
\,A(\tau )\,|i,n} \right\rangle }
\end{equation} 
where 
  $A(0)=\rho _{SS}\,a^\dagger$ and $ \dot A={\cal L}A$.
The resulting equations can be written in the form
\begin{equation}
{{d\vec A } \over {dt}}=\mathord{\buildrel{\lower3pt\hbox{$
\scriptscriptstyle\leftrightarrow$}}\over M}\vec A 
\end{equation}
with 
\begin{equation}
\vec A= 
\left(
\begin{array}{c}
A_{0,-;0,+} \\
A_{0,-;1,-}\\
A_{1,-;0,-}\\
A_{1,+;0,+}\\
A_{2,-;1,-}\\
A_{0,+;0,-}\\
A_{1,+;1,-}\\
A_{2,-;0,+}
\end{array}
\right)
\end{equation}
with the notation $A_{n,\pm ;m,\pm} \equiv \langle n,\pm | A | m,\pm 
\rangle$ and initial conditions 
\begin{equation}
A_{n,\pm ;m,\pm} (0)= \langle n,\pm | a^\dagger \rho_{ss} | m,\pm \rangle = 
\sqrt{n} \langle n-1,\pm | \rho_{ss} | m,\pm \rangle\end{equation}
The matrix $M$ is given as
\begin{equation}
M= 
\left(
\begin{array}{cccccccc}
-\gamma /2 &-g &0&0&0&0&0&0\\
g&-\kappa &0&0&0&0&0&0 \\
0&\sqrt{2}F&-\kappa & \gamma & 2\sqrt{2}\kappa & g & 0 & 0   \\
0&0&0&-(\gamma +\kappa) & 0 & 0  & -g & -\sqrt{2}g \\
0 & 0 & 0 &0& -3\kappa & 0&\sqrt{2}g & g \\
0 & 0 & -g &0 & 0 &-\gamma /2 &2\kappa & 0\\
0&0&0&g&-\sqrt{2}g&0&-(\gamma /2+2\kappa) &0\\
\sqrt{2}F&0&0&\sqrt{2}g&-g&0&0 &-(\gamma/2 +2\kappa)
\end{array}
\right)
\end{equation}
After taking the Fourier transform of the above equations we have
\begin{equation}
 \vec {\tilde A}(\omega )={\left\{ \mathord{\buildrel{\lower3pt\hbox{$
 \scriptscriptstyle\leftrightarrow$}}\over M}-i\omega
\mathord{\buildrel{\lower3pt\hbox{$\scriptscriptstyle\leftrightarrow$}}
\over I} \right\}}^{-1} \vec A (0)
\end{equation}
with $\vec {\tilde A}(\omega )$ composed of the Fourier transform of $
\vec A (\tau)$ and then we can easily form the spectrum
\begin{equation}
I_{tr}(\omega )=\sum\limits_{i,n} {\sqrt {n+1}\,\left\langle {i,n+1\,|\,
\Re \;\tilde A(\omega )\,|i,n} \right\rangle }
\end{equation}
We will also be interested in the spectrum of squeezing, defined as 
\begin{equation}
S(\omega ,\theta )=\int^{\infty}_{-\infty} d\tau \cos \omega \tau Re\left[
\langle \Delta a^\dagger (\tau )\Delta a(0)\rangle +e^{2i\theta } \langle 
\Delta a^\dagger (\tau )\Delta a^\dagger (0)\rangle \right], \label{seq}
\end{equation} 

Adding two spectra of squeezing, with phase angles $\theta $ and
$\theta +\pi /2 $, we obtain the following relationship
between the incoherent spectrum and the spectrum of squeezing,
\begin{equation}
I_{inc} (\omega) \propto \left[S(\omega ,\theta )+
S(\omega ,\theta +\pi /2)\right]
\end{equation}
For fields whose fluctuations can be described by a classical
stochastic process, both
$S(\omega ,\theta )$ and
$S(\omega ,\theta +\pi /2)$
must be positive.  As noted above, for a squeezed quantum field, one of
these spectra is negative over some range of frequencies, for
appropriate choice of the phase $\theta $.
To calculate the second term in equation \ref{seq} we must use the quantum 
regression theorem for
 \begin{equation}
\left\langle {\Delta a^\dagger (0) \Delta a^\dagger\,(\tau )} \right\rangle 
=tr\left\{ {a^\dagger (0)\,B(\tau )} \right\}=\sum\limits_{i,n} {\sqrt {n}\,
\left\langle {i,n-1\,|\,B(\tau )\,|i,n} \right\rangle }
\end{equation} 
where 
  $B(0)=\rho _{SS}\,a^\dagger$ and $ \dot B={\cal L}B$ and the nonzero 
  elements of interest are
\begin{equation}
\vec B= 
\left(
\begin{array}{c}
B_{0,-;0,+} \\
B_{0,-;1,-}\\
B_{1,+;0,+}\\
B_{2,-;1,-}\\
B_{1,+;1,-}\\
B_{2,-;0,+}\\
B_{0,+;0,-}\\
B_{1,-;0,-}
\end{array}
\right)
\end{equation}
The relavant equations are
\begin{equation}
{{d\vec B } \over {dt}}=\mathord{\buildrel{\lower3pt\hbox{$
\scriptscriptstyle\leftrightarrow$}}\over M}\vec B 
\end{equation}

The second term dominates, and is proportional to $F$. The spectrum of 
squeezing that is plotted is $S(\omega,0)$. The spectrum of squeezing for the 
quadrature $\pi/2$ out of phase, $S(\omega,\pi/2)$ is equal and opposite in 
sign.
 In the system under consideration here we find that
\begin{equation}
S(\omega ,\theta )=-
S(\omega ,\theta +\pi /2)
\end{equation}
To first order in $F$, these two contributions to the incoherent spectrum 
cancel, but they differ in terms of order $F^2$. This means that the 
incoherent spectrum is formed by the
subtraction of two quantities, in the presence of squeezing. 

We now turn to results for the incoherent spectrum. In Figure 2-8, we plot 
the incoherent spectrum and spectrum of squeezing for $\kappa/\gamma =10.0$ 
and various values of atom-field coupling $g$. In all Figures, the solid line 
is the incoherent spectrum, and the dotted line is the spectrum of squeezing 
for the quadrature in phase with the driving field. In Figure 2, for 
$g/\gamma=0.1$, the spectrum is essentially a Lorentzian with linewidth 
$\kappa (1+2g^2/\kappa \gamma)$. In Figure 3, with $g/\gamma =0.3$, a hole 
appears in the spectrum, which deepens with further increases in $g$. In 
Figure 4 where $g/\gamma=3.0$, a small bump appears inside the hole. 
Increasing $g/\gamma$ to $5.0$ leads to the bump inside the hole increasing 
in size, as in Figure 5. As $g/\gamma$ is increased to  $g/\gamma=10.0$, in 
Figure 6, the spectrum appears to have a double dip in it. These dips appear 
near $\omega =\pm g$. We note that this is not a hole due to absorption of 
energy emitted out the side of the cavity, as it persists in the limit 
$\gamma \rightarrow 0$. As one increases $g/\gamma$ to  $15$ a double peaked 
structure appears as in vacuum-Rabi splitting, as shown in Figure 7. 
Increasing $g$ to $50$ leads to a well defined vacuum-Rabi structure. Each 
vacuum-Rabi peak has a hole in it however. These holes are deepened if we 
decrease $\gamma$ relative to $\kappa$ and $g$ as shown in Figure 8. Again 
this points out that these holes are not just due to fluorescence out of the 
side of the cavity. In the good cavity limit,  $\kappa /\gamma \ll 1$, we find 
for small $g/\gamma$ a subnatural width single peaked spectrum (Figure 9), 
which evolves into a vacuum-Rabi doublet with no holes  for large $g/\gamma$ 
(Figure 10). Hence in the good cavity limit, the anomalous effects disappear.

\section{Optical Spectrum of the Fluorescent Light}

\begin{eqnarray}
I_{fl}(\omega )&=& \int\limits_{-\infty }^\infty  {d\tau \,e^{i\omega \,\tau 
}}\left\langle {\sigma_+(0)\,\sigma_-\,(\tau )} \right\rangle 
\nonumber \\
&=&2\Re \int\limits_0^\infty  {d\tau \,e^{i\omega \,\tau }}\left\langle 
{\sigma_+(0)\,\sigma_-\,(\tau )} \right\rangle \nonumber \\
&=&2\pi \langle \sigma_+\rangle_{ss} \langle \sigma_- \rangle_{ss} \delta 
(\omega) +\nonumber \\
&&2\Re \int\limits_0^\infty  {d\tau \,e^{i\omega \,\tau }}\left\langle 
{\Delta \sigma_+(0)\,\Delta \sigma_-\,(\tau )} \right\rangle
\end{eqnarray}

Again here there is no coherent, or elastic scattering leading to a delta 
function component of the spectrum at resonance.

By the quantum regression theorem we have
\begin{equation}
\left\langle {\sigma_+\,\sigma_-\,(\tau )} \right\rangle =tr\left\{ {
\sigma_-(0)\,C(\tau )} \right\}=\sum\limits_{n} {\,\left\langle {+,n\,|\,C(
\tau )\,|-,n} \right\rangle }
\end{equation} 
where 
  $C(0)=\rho _{SS}\,\sigma_+$ and $ \dot C={\cal L}A$
Hence we can write the incoherent spectrum as
\begin{equation}
I_{fl}(\omega )=\sum\limits_{i,n} {\sqrt {n+1}\,\left\langle {i,n+1\,|\,\Re 
\;\tilde A(\omega )\,|i,n} \right\rangle }
\end{equation}

The resulting equations can be written in the form
\begin{equation}
{{d\vec C } \over {dt}}=\mathord{\buildrel{\lower3pt\hbox{$
\scriptscriptstyle\leftrightarrow$}}\over M}\vec C 
\end{equation}
with 

\begin{equation}
\vec C= 
\left(
\begin{array}{c}
C_{0,-;0,+} \\
C_{0,-;1,-}\\
C_{0,+;0,-}\\
C_{1,+;1,-}\\
C_{1,-;0,-}\\
C_{1,+;0,+}\\
C_{2,-;1,-}\\
C_{2,-;0,+}
\end{array}
\right)
\end{equation}
with the notation $C_{n,\pm ;m,\pm} \equiv \langle n,\pm |C | m,\pm \rangle$ 
and initial conditions 
\begin{eqnarray}
C_{n,- ;m,\pm} (0)&= &\langle n,\pm | \sigma_+ \rho_{ss} | m,\pm \rangle = 
\langle n,+ | \rho_{ss} | m,\pm \rangle\\
C_{n,+ ;m,\pm}&=&0
\end{eqnarray}

After taking the Fourier transform of the above equations,  we have

\begin{equation}
 \vec {\tilde C}(\omega )={\left\{ \mathord{\buildrel{\lower3pt\hbox{$
 \scriptscriptstyle\leftrightarrow$}}\over M}-i\omega
\mathord{\buildrel{\lower3pt\hbox{$\scriptscriptstyle\leftrightarrow$}}
\over I} \right\}}^{-1} \vec C (0)
\end{equation}
with $\vec {\tilde C}(\omega )$ composed of the Fourier transform of $
\vec C (\tau)$ and then we can easily form the fluorescent spectrum
\begin{equation}
S(\omega )=\sum\limits_{n} \langle +,n|\Re \tilde C(\omega )\,|-,n \rangle
\end{equation}

As before, we will be interested in the spectrum of squeezing of the 
fluorescent light.
\begin{equation}
S_{fl}(\omega ,\theta )=\int^{\infty}_{-\infty} d\tau \cos \omega \tau Re
\left[
\langle \Delta \sigma_+ (\tau )\Delta \sigma_-(0)\rangle +e^{2i\theta } 
\langle \Delta \sigma_+ (\tau )\Delta \sigma_+ (0)\rangle \right],
\end{equation} 

To calculate the second term in the above equation, we must use the quantum 
regression theorem for
 \begin{equation}
\left\langle {\Delta \sigma_+ (0) \Delta \sigma_+\,(\tau )} \right\rangle 
=tr\left\{ {\sigma_+ (0)\,D(\tau )} \right\}=\sum\limits_{n} \,\left\langle 
{+,n\,|\,D(\tau )\,|-,n} \right\rangle 
\end{equation} 
where 
  $D(0)=\rho _{SS}\,\sigma_+$ and $ \dot D={\cal L}D$ and the nonzero elements 
  of interest are
\begin{equation}
\vec D= 
\left(
\begin{array}{c}
D_{0,-;0,+} \\
D_{0,-;1,-}\\
D_{1,+;0,+}\\
D_{2,-;1,-}\\
D_{1,+;1,-}\\
D_{2,-;0,+}\\
D_{0,+;0,-}\\
D_{1,-;0,-}
\end{array}
\right)
\end{equation}
The relevant equations are
\begin{equation}
{{d\vec D } \over {dt}}=\mathord{\buildrel{\lower3pt\hbox{$
\scriptscriptstyle\leftrightarrow$}}\over M}\vec D 
\end{equation}

In the case of the fluorescent light, in the bad cavity limit  ( $\kappa 
/\gamma \gg 1$), we have a single peaked structure with no holes for 
$g/\gamma \ll 1$, as in Figure 11.
 Keeping $\kappa /\gamma \gg 1$, and with $g/\gamma \gg 1$, we have a 
 vacuum-Rabi doublet with no holes as in Figure 12. So for the flourescent 
 light there are no anomalous effects in the spectra in the bad cavity limit. 
 In Figure 13,  we let $\kappa /\gamma \ll 1$, and there are no holes for  
 $g/\gamma \ll 1$. For $\kappa /\gamma \ll 1$, and with $g/\gamma \gg 1$ we 
 see a vacuum-Rabi doublet with holes in Figure 14. The holes are deepened as 
 one goes further into the good cavity limit as shown in Figure 15. So it is 
 in the good cavity limit that we see anomalous effects in the spectra for 
 the fluorescent light from this system.

\section {Physical Interpretation}
So far we have seen incoherent spectra with a subnatural linewidth, and also 
ones with spectral holes. Similar types of spectra have been predicted for a 
single two-level atom in a microcavity driven by a weak external field 
resonant with the atom and the cavity \cite{HJCPRR}. That is essentially the 
system we consider in the present paper, but without the $\chi^{(2)}$ 
crystal, and driven at $\omega_0$, and not $2\omega_0$. In that case the 
spectrum of squeezing was proportional to $E^2$, where $E$ is the strength of 
the driving field at $\omega_0$, and the incoherent spectrum is proportional 
to $E^4$. In particular, in the bad-cavity limit of the previous system, where 
$\kappa \gg g, \gamma$, it was found there that the incoherent spectrum of the 
transmitted light was a Lorentzian squared. The Lorentzian had a linewidth of 
$\delta =\gamma (1+2g^2/\kappa \gamma)$, which is the cavity enhanced 
spontaneous emission rate. As the spectrum is the square of that Lorentzian, 
the linewidth is about $\Delta \omega \approx 0.67 \delta$. This result also 
obtains for the incoherent spectrum of a driven two-level atom in free space, 
i.e. resonance fluorescence. There, for weak driving fields $\delta =\gamma$, 
and a subnatural linewidth results from the squared Lorentzian, as first noted 
by Mollow \cite{Mollow}. For that same system, in the strong coupling limit, 
for $g \gg \kappa , \gamma$, a vacuum-Rabi doublet was found, each peak being 
a squared Lorentzian with $\delta= (2\kappa +\gamma)/2$. In the good cavity 
limit of that driven atom-cavity system, $\kappa \ll g \ll \gamma$, a single 
peaked structure with a hole appeared as the incoherent spectrum. Again the 
depth of that hole reached zero as $\gamma \rightarrow 0$, and so does not 
represent a loss of photons at line center due to absorption and emission out 
the side. These phenomena were referred to as squeezing induced linewidth 
narrowing (SILN) and squeezing induced spectral holes (SISH). Recall that the 
incoherent spectrum is the sum of two squeezing spectra $\pi /2$ out of phase 
with one another. In the case of resonance fluorescence, the two spectra of 
squeezing were both single peaked functions, and were equal and opposite to 
order $E^2$. Keeping terms to order $E^4$, we found there that the two 
spectra of squeezing were both Lorentzians, but one was negative (indicating 
squeezing) and the other positive. Hence the Lorentzian squared was formed 
from the subtraction of two Lorentzians, one with a linewidth of $\gamma /2$ 
from which is subtracted one with a larger width. This is shown schematically 
in Figure 13. Spectral holes were shown in \cite{HJCPRR} to arise in a similar 
manner, when the incoherent spectrum is the subtraction of two Lorentzian-like 
structures which are equal at line center but differ in the wings. These holes 
are only nonclassical if the two squeezing spectra are single peaked 
structures, as discussed in \cite{HJCPRR}. 

As the subtraction results from one of the spectra of squeezing being 
negative, indicating fluctuations in that quadrature below the vacuum noise 
level, it was inferred that these narrowings and holes result from the fact 
the the light emitted in the incoherent spectrum is squeezed. At the time, 
theoretical investigations had shown that shining squeezed light on an optical 
system could reduce the effective linewidth of spontaneous emission from that 
system, by altering the vacuum fluctuations that the unstable excited state 
coupled to. So it was proposed that the narrowings/holes seen in the 
incoherent spectra resulted from the fact that the radiation reaction force on 
the optical system was squeezed, instead of the vacuum fluctuations. The 
amount of squeezing is vanishingly small in the weak field limit, and in 
retrospect it seems odd that a vanishingly small amount of squeezing could 
result in a $33\%$ reduction in linewidth. Further, the effects of spectral 
holes and narrowings go away as the driving field strength is increased, and 
the amount of squeezing increases. Another example of this is the optical 
parametric oscillator (OPO). The output spectrum of that device is a 
Lorentzian squared for weak pumping fields, with $\delta = \kappa$. The OPO 
produces a vanishingly small amount of squeezing in that limit. It is a good 
source of squeezed light at higher pump fields, with large amounts of 
squeezing produced just below the oscillation threshold. But the linewidth is 
not narrow in that instance.  From Figures 2-8, we see that in the system 
under consideration here the physics is probably more complicated. The 
spectra of squeezing are complex structures that do not yield themselves to 
the type of interpretation suggested in \cite{HJCPRR}.

We now consider another possible mechanism for holes and  narrowing to appear 
in incoherent spectra. Recall that the incoherent spectrum results from a 
nonlinear scattering process involving two or more photons. The effect is 
most evident for weak driving fields, where two photons are emitted from the 
$\chi^{(2)}$ crystal, and interact with the atom-cavity system. After several 
cavity and/or spontaneous emission lifetimes, the interaction is completed by 
the emission of two photons. This can happen via emission of two photons into 
the cavity mode, one into the cavity mode and one out the side of the cavity, 
or both out the side of the cavity.  In the weak field limit $F \ll g, 
\kappa,$ and $\gamma$, the next pair of photons from the nonlinear crystal 
arise long after the previous two-photon scattering process is completed. The 
two emitted photons are highly correlated, as their frequencies must satisfy 
energy conservation $\omega_1 +\omega_2 =2\omega_0$, which requires that the 
two photons be emitted at frequencies $\omega_0 \pm \delta \omega$. The 
emitted photons momenta must similarly satisfy conservation of momentum as 
$\vec k_1 +\vec k_2 = 2\vec k_0$. Single photon scattering events lead to the 
delta function component of the spectrum, the elastic, or coherent scattering. 
There is no contribution to that in our system, but there may be in other 
nonlinear optical systems. So the thought occurs that perhaps the root cause 
of the anomalous effects (holes and narrowings)  are due to quantum 
interference between various indistinguishable emission pathways, akin to 
similar effects in absorption (e.g. electromagnetically induced transparency) 
and spontaneous emission from a given initial unstable state. 

In the case of resonance fluorescence, it has recently been shown 
\cite{Pedrotti} that the Lorentzian squared results from quantum interference, 
as the probability to obtain a photon in mode $k$ can be written (in the weak 
field limit) as
\begin{equation}
\left|  c_{k}\right|  ^{2}=\left|  c_{b1k}+\sum_{k^{\prime}}c_{b1k1k^{
\prime}%
}+\sum_{k^{\prime}}c_{b1k^{\prime}1k}\right|  ^{2}%
\label{ceq}
\end{equation}
The first term, $c_{b1k}$ is the probability amplitude for scattering one 
photon into mode $k$, the coherent scattering. This piece gives rise to a 
delta function, so-called coherent spike, to the spectrum. The other two terms 
are two-photon scattering terms, where a pair of photons has been scattered, 
one into mode $k$, and one into mode $k'$. There are two ways for the 
two-photon scattering to happen. The $k$ photon may come before or after the 
$k'$ photon, as represented by $c_{b1k1k^{\prime}}$ and $c_{b1k^{\prime}1k}$. 
If these two probability amplitudes are nonzero over an overlapping range of 
$k'$ modes, then there is a cross term that results in equation \ref{ceq}. We 
propose that this mechanism is also responsible for the anomalous spectra 
observed in the case of the two level atom inside a weakly driven optical 
parametric oscillator, and indeed for all weakly driven nonlinear optical 
systems. It is instructive to look at quantum trajectories for this system. 
In this case we describe the system by a conditioned wave function, a 
non-Hermitian Hamiltonian, and associated collapse processes. These are given 
by
\begin {eqnarray}
|\psi _c(t)\rangle &=&\sum\limits_{n=0}^\infty  {\alpha_{g,n}(t)e^{-i
E_{g,n}t}|g,n\rangle +\alpha_{e,n}(t)e^{-iE_{e,n}t}|e,n\rangle }\\
H_D&=& -i\kappa a^\dagger a + -i\gamma /2 \sigma_+ +i\hbar F(a^{\dagger ^2}
-a^2)+i\hbar g\;(a^\dagger \sigma _--a\sigma _+)
\end{eqnarray}
where we also have collapse operators  $C_{cav}=\sqrt{\kappa} a$ and 
$C_{spon. em.}=\sqrt{\gamma /2}\sigma_-$.

In Figure 16, we plot $\langle \psi_{cond} \mid a^\dagger a \mid \psi_{cond} 
\rangle$ as a function of time, in a case where $\gamma \ll \kappa$. The 
system is in steady state, and then a photon emission occurs out the front of 
the cavity. The conditioned photon number rises to unity \cite{HJCbook}. This 
is because we know that photons are created in pairs in this system, and 
detection of one outside the cavity means that one must remain. We know that 
the first photon detected is at $\omega_0 +\delta \omega$, and that the photon 
that remains  inside the cavity is of frequency $\omega_0-\delta \omega$. 
However we are unsure what the value of $\delta \omega$ is. In particular, is 
$\delta \omega$ greater than or less than zero? In other words, is the first 
emission event the photon that falls to the right or left of the resonant 
frequency in the incoherent spectra? It is this indistinguishability that 
leads to the spectra we present. We should expect different results for the 
fluorescent spectrum in this case. When $\gamma \ll \kappa$, it is most likely 
that the two photons will exit the system through the cavity mirror. 
Occasionally, one leaves via the cavity mirror and one is emitted out the side 
of the cavity. Even more rare in this limit is two photons scattered out the 
side of the cavity. There is no narrowing or hole in the fluorescent spectrum. 
This is because there is no quantum interference in this case. The photon 
detected in fluorescence is most probably associated with another photon 
emitted out the cavity mirror. These photons are distinguishable in the sense 
that we know which direction they have been emitted into, and hence no 
interference. We see a similar type of thing in the limit where $\gamma \gg 
\kappa$, where we see anomalous spectra in fluorescence (where pairs of 
photons are most likely emitted) and not in transmission (where a transmitted 
photon is most likely paired with a fluorescent photon). This lends credence 
to our proposal that quantum interference is responsible for the spectral 
narrowing and holes.

At higher driving field strengths, there are more terms in equation \ref{ceq} 
, which are added and then squared to get the probability of obtaining the 
photon at a given $k$. The relative phase of the complex amplitudes is such 
that the size of the cross, or ``interference" terms becomes smaller.  If for 
example, we have two two-photon scattering events within a cavity lifetime. 
Two photons detected in the output of the cavity may or may not have been 
correlated before they were scattered. This type of behavior can be seen in 
Figure 17, were we plot $\langle \psi_{cond} \mid a^\dagger a \mid \psi_{cond} 
\rangle$ as a function of time for larger driving fields. This will tend to 
reduce the size of the effect. Inasmuch as we are considering a system driven 
by very weakly squeezed light, if one drives an optical system with a weakly 
squeezed field with no coherent component, or a weakly squeezed vacuum, 
similar effects should be obtained. This is indeed the case as shown by the 
work of Swain et. al. \cite{Swain,Swain1} We then conclude that these types of 
anomalous spectra in weakly driven nonlinear optical systems is indeed due to 
the type of quantum interference ala equation \ref{ceq}.

\section{Conclusion}

We have shown that the transmitted and fluorescent incoherent spectra of a 
two-level atom in a weakly driven optical parametric oscillator can exhibit 
spectral holes and spectral narrowing. These types of phenomena have been 
predicted for other nonlinear optical systems, but the previous description of 
why they occur has been found lacking. We propose a new mechanism for these 
effects, based on recent work on resonance fluorescence. Further work on this 
system and others should lead to a better understanding of such anomalous 
spectra, and indeed the difference between a spontaneous emission spectrum 
for a system prepared in a particular unstable state, and the driven type of 
spectra that we consider here.

We would like to thank Dr.  Leno Pedrotti, Dr. Min Xiao, Dr. Julio 
Gea-Banacloche, and Dr. Tom Mossberg for helpful conversations.

\begin{figure}
\caption{A schematic of the physical system under consideration. We have a 
single two-level atom in a resonant cavity. $F(2\omega)$ is a classical 
driving field at twice the resonant frequency. The nonlinear crystal has a 
second order susceptibility $\chi^{(2)}$. $g$ is the atom-field coupling, 
$\kappa$ is the field decay rate through the righthand mirror, and $\gamma$ 
is the spontaneous emission rate to non-cavity modes.}
\end{figure}

\begin{figure}
\caption{ Spectrum of the transmitted light for $\kappa /\gamma=10.0$, and 
$g/\gamma=0.1$. The dotted line is the spectrum of squeezing for the in-phase 
quadrature.}
\end{figure}

\begin{figure}
\caption{ Spectrum of the transmitted light for $\kappa /\gamma=10.0$, and 
$g/\gamma=1.0$. The dotted line is the spectrum of squeezing for the in-phase 
quadrature.}
\end{figure}

\begin{figure}
\caption{ Spectrum of the transmitted light for $\kappa /\gamma=10.0$, and 
$g/\gamma=3.0$. The dotted line is the spectrum of squeezing for the in-phase 
quadrature.}
\end{figure}

\begin{figure}
\caption{ Spectrum of the transmitted light for $\kappa /\gamma=10.0$, and 
$g/\gamma=5.0$. The dotted line is the spectrum of squeezing for the in-phase 
quadrature.}
\end{figure}

\begin{figure}
\caption{ Spectrum of the transmitted light for $\kappa /\gamma=10.0$, and 
$g/\gamma=10.0$. The dotted line is the spectrum of squeezing for the in-phase 
quadrature.}
\end{figure}

\begin{figure}
\caption{ Spectrum of the transmitted light for $\kappa /\gamma=10.0$, and 
$g/\gamma=50.0$. The dotted line is the spectrum of squeezing for the in-phase 
quadrature.}
\end{figure}

\begin{figure}
\caption{ Spectrum of the transmitted light for $\kappa /\gamma=100.0$, and 
$g/\gamma=30.0$. The dotted line is the spectrum of squeezing for the in-phase 
quadrature.}
\end{figure}

\begin{figure}
\caption{ Spectrum of the transmitted light for $\kappa /\gamma=0.1$, and 
$g/\gamma=0.1$. The dotted line is the spectrum of squeezing for the in-phase 
quadrature.}
\end{figure}

\begin{figure}
\caption{ Spectrum of the transmitted light for $\kappa /\gamma=0.1$, and 
$g/\gamma=20.0$. The dotted line is the spectrum of squeezing for the in-phase 
quadrature.}
\end{figure}

\begin{figure}
\caption{ Spectrum of the fluorescent light for $\kappa /\gamma=10.0$, and 
$g/\gamma=3.0$. The dotted line is the spectrum of squeezing (scaled down by 
a factor of $10$) for the in-phase quadrature.}
\end{figure}

\begin{figure}
\caption{ Spectrum of the fluorescent light for $\kappa /\gamma=10.0$, and 
$g/\gamma=50.0$. The dotted line is the spectrum of squeezing (scaled down by 
a factor of $10$) for the in-phase quadrature.}
\end{figure}

\begin{figure}
\caption{ Spectrum of the fluorescent light for $\kappa /\gamma=0.1$, and 
$g/\gamma=0.3$. The dotted line is the spectrum of squeezing (scaled down by 
a factor of $10$) for the in-phase quadrature.}
\end{figure}

\begin{figure}
\caption{ Spectrum of the fluorescent light for $\kappa /\gamma=0.1$, and 
$g/\gamma=10.0$. The dotted line is the spectrum of squeezing (scaled down by 
a factor of $10$) for the in-phase quadrature.}
\end{figure}

\begin{figure}
\caption{ Spectrum of the fluorescent light for $\kappa /\gamma=0.01$, and 
$g/\gamma=5.0$. The dotted line is the spectrum of squeezing (scaled down by 
a factor of $10$) for the in-phase quadrature.}
\end{figure}

\begin{figure}
\caption{ Conditioned mean intracavity photon number for $g/\gamma=1.0$, 
$\kappa /\gamma=10.0$, and $F/\gamma=0.1$}
\end{figure}

\begin{figure}
\caption{Conditioned mean intracavity photon number for $g/\gamma=40.0$, 
$\kappa /\gamma=10.0$, and $F/\gamma=1.0$}
\end{figure}

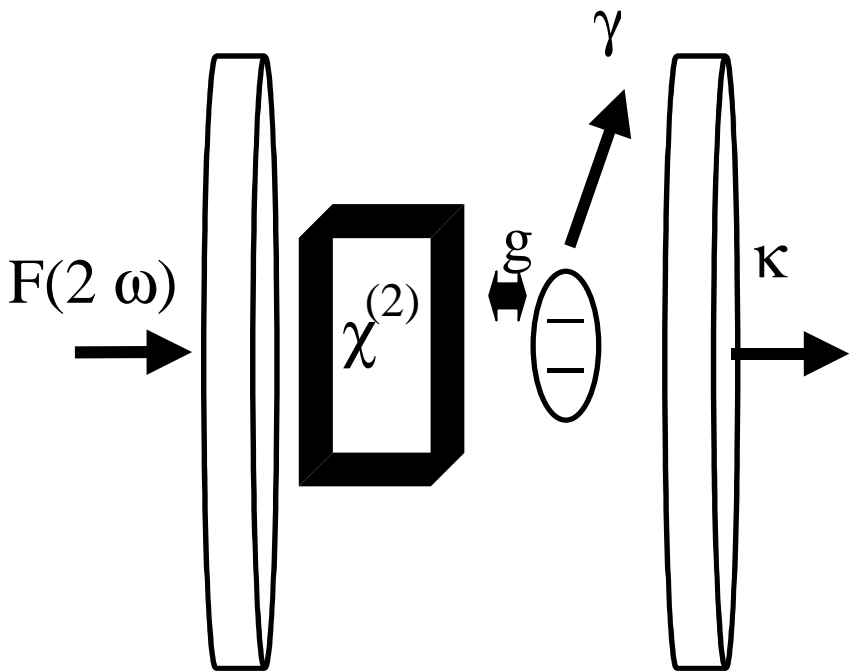
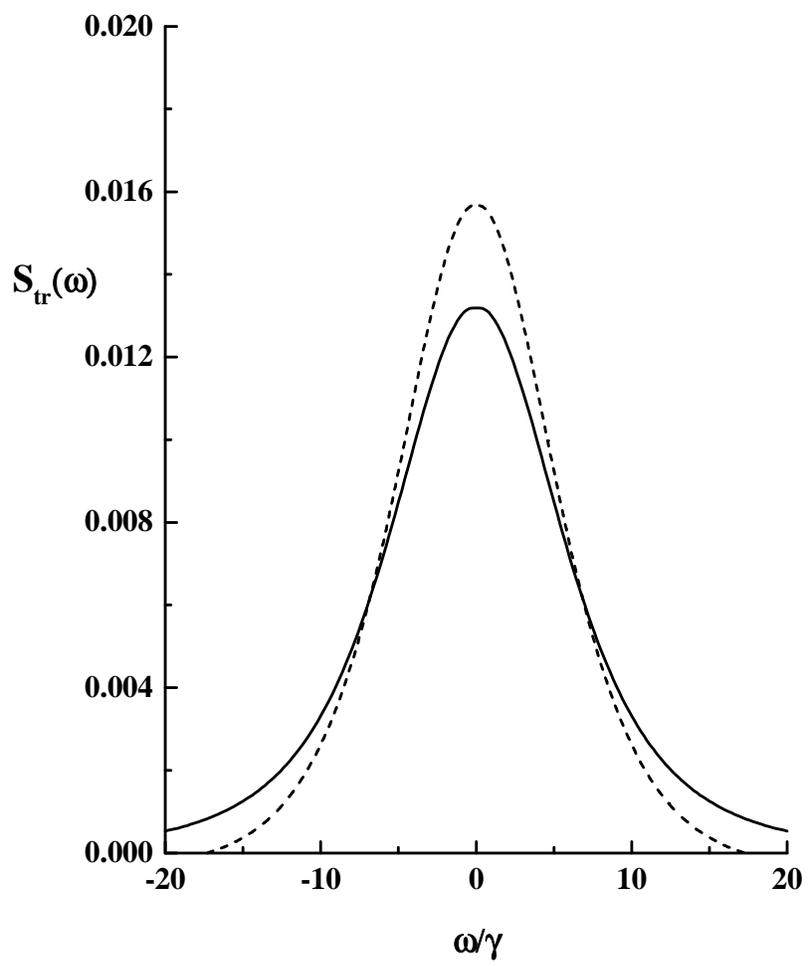
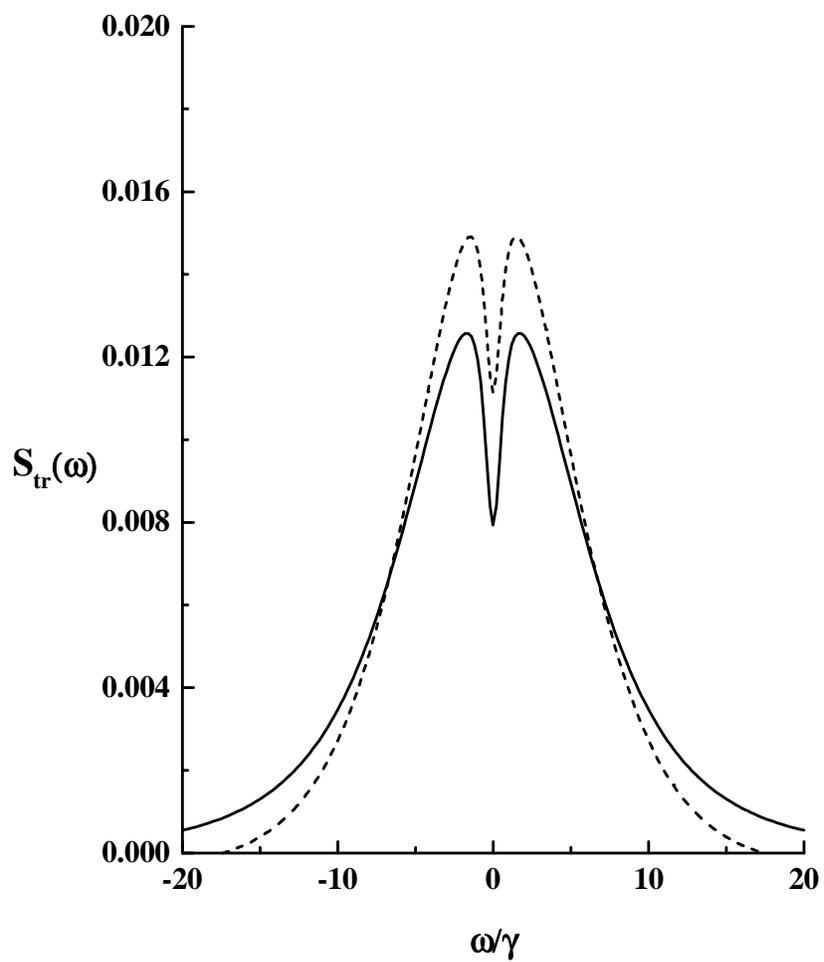
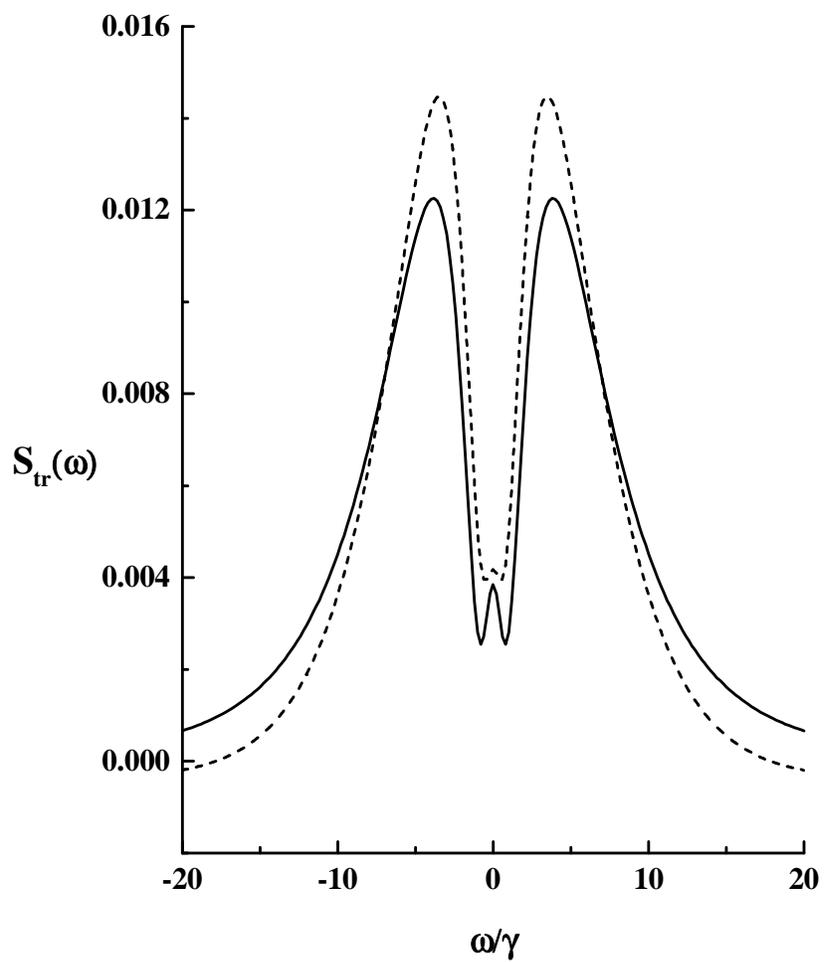
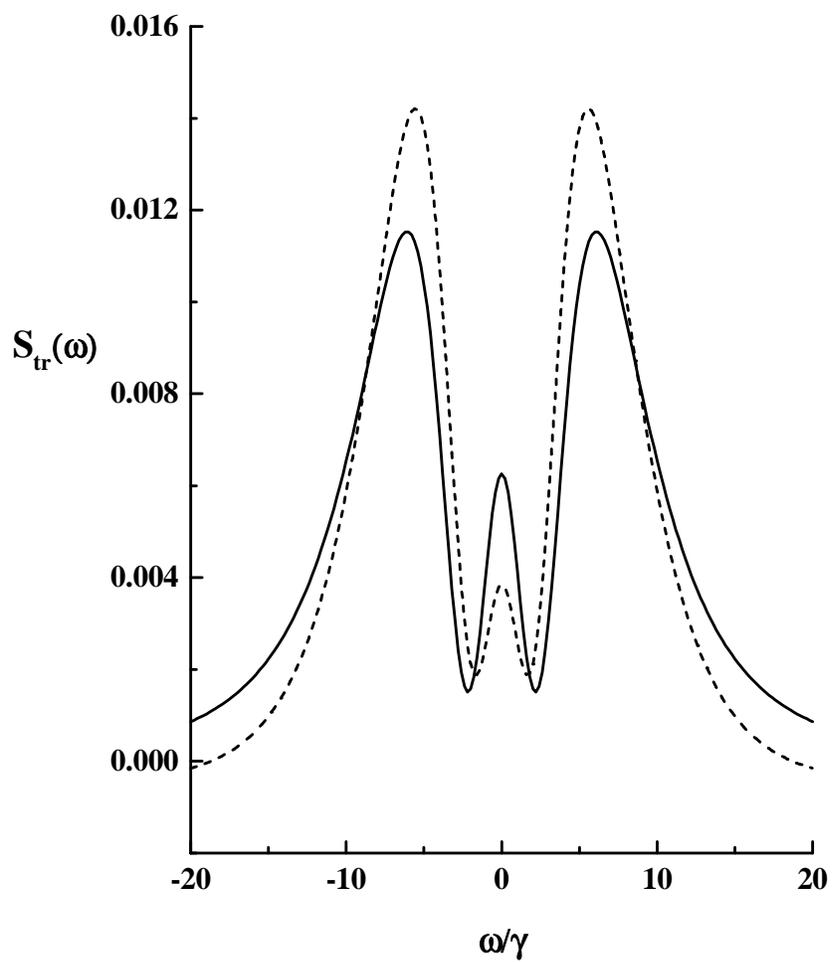
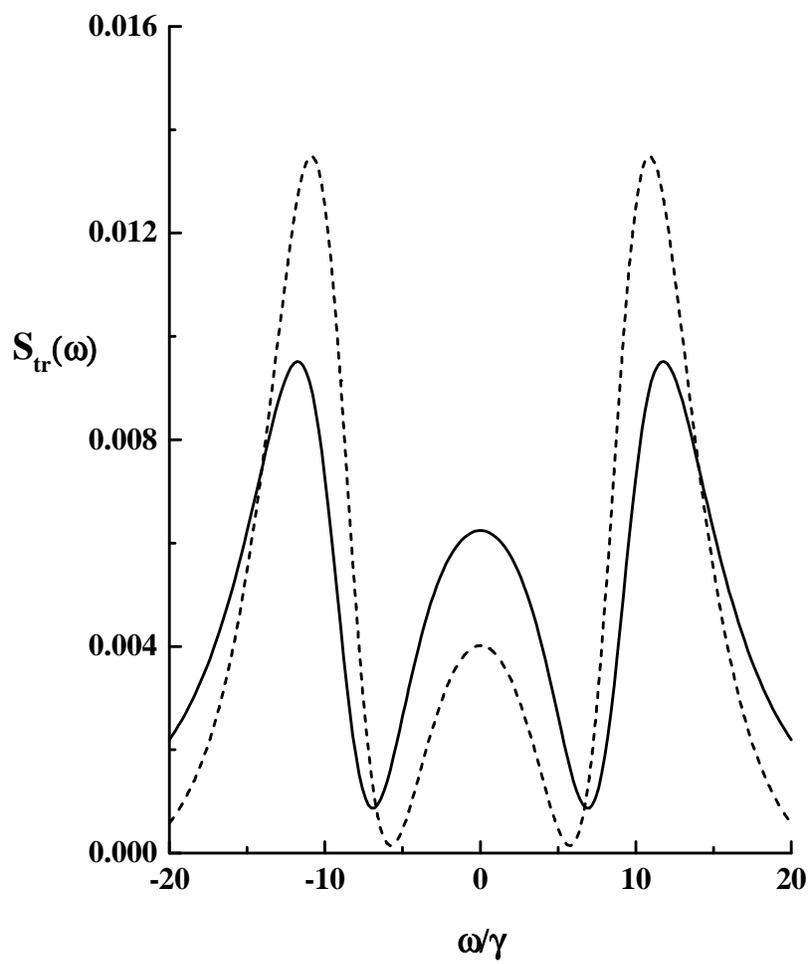
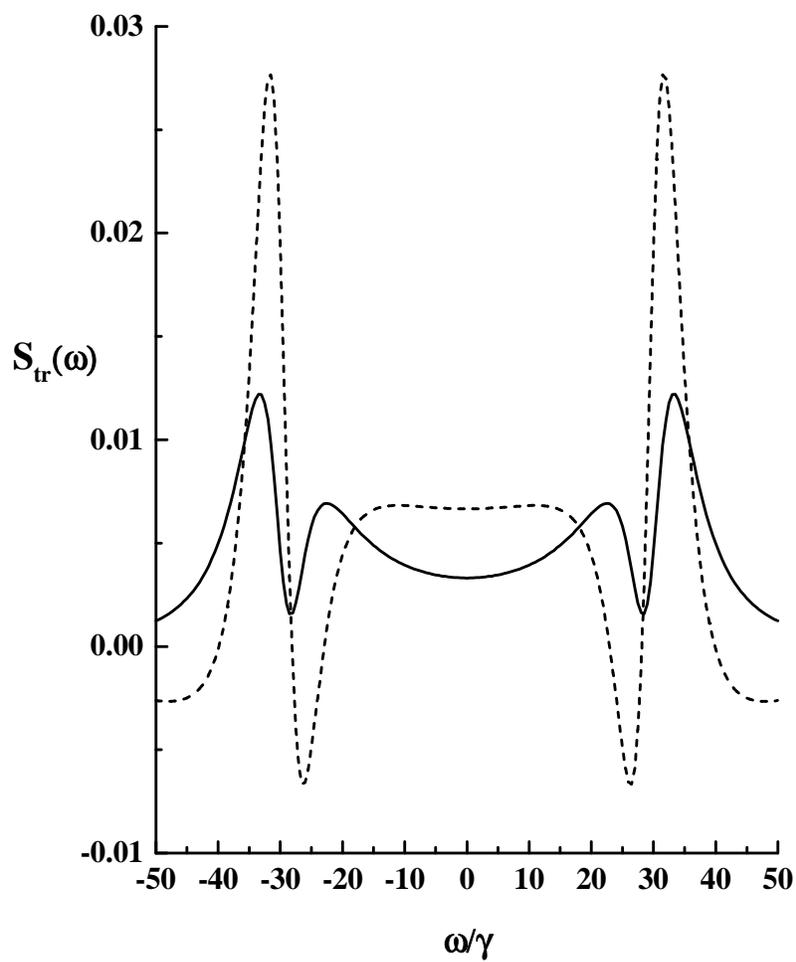
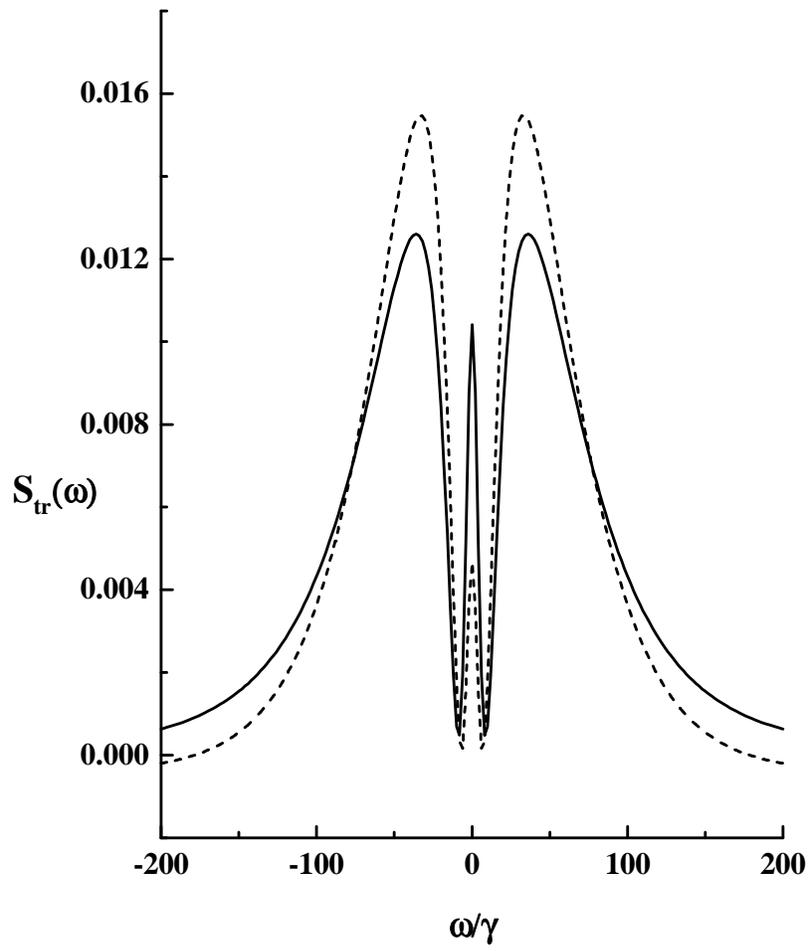
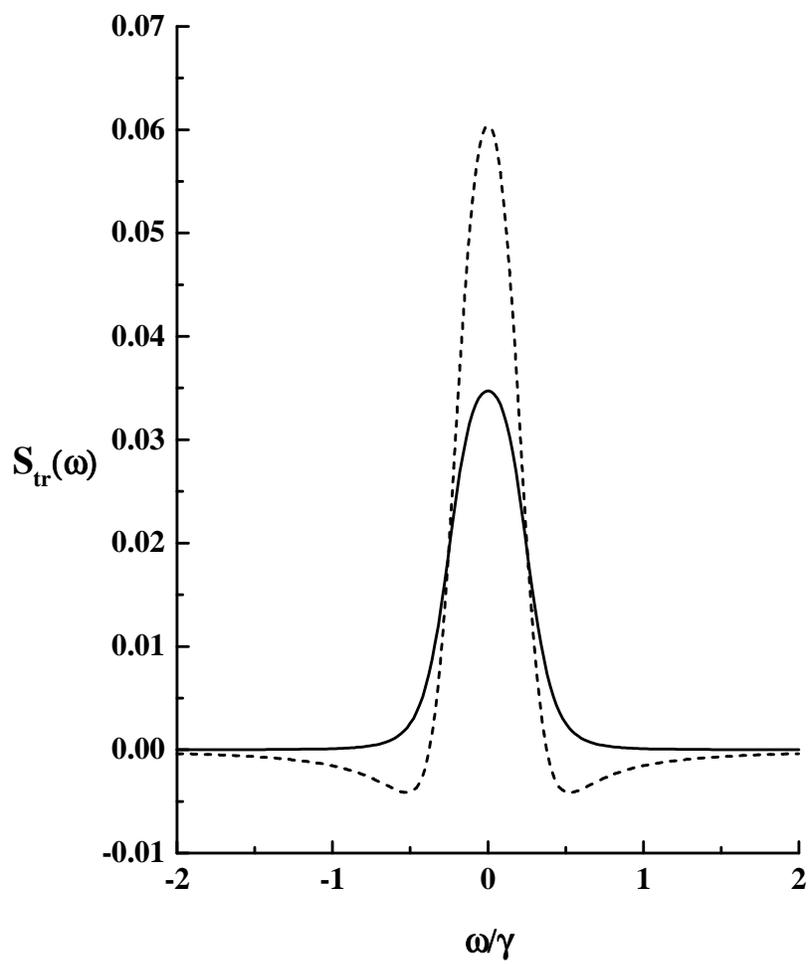
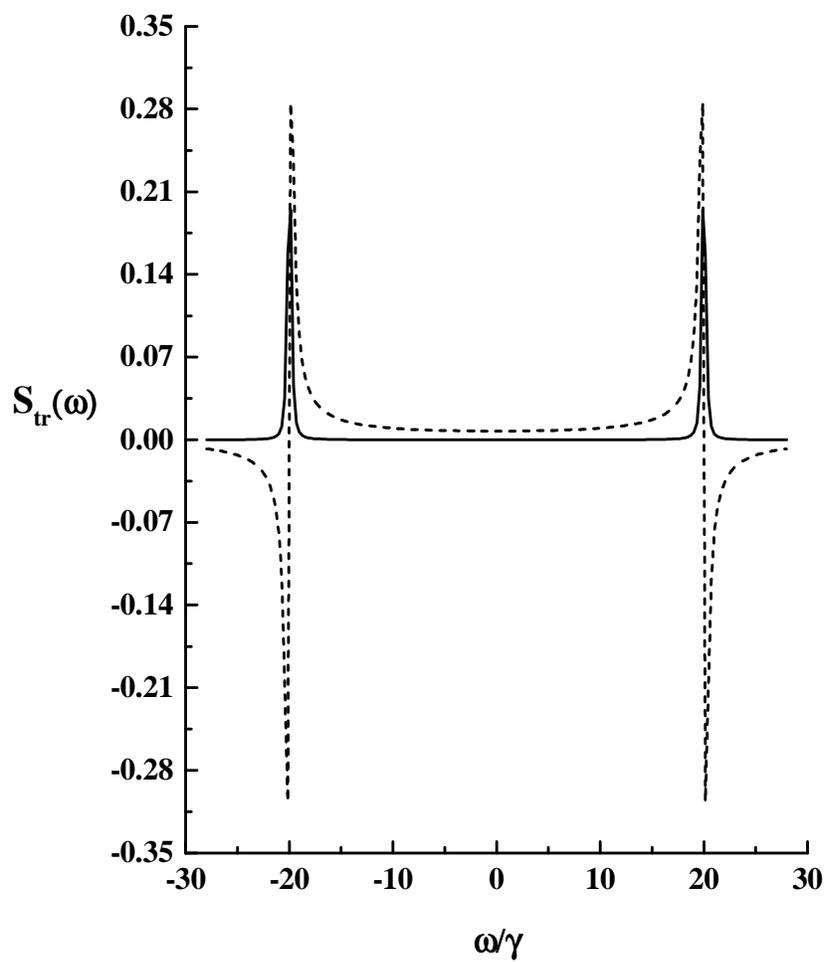
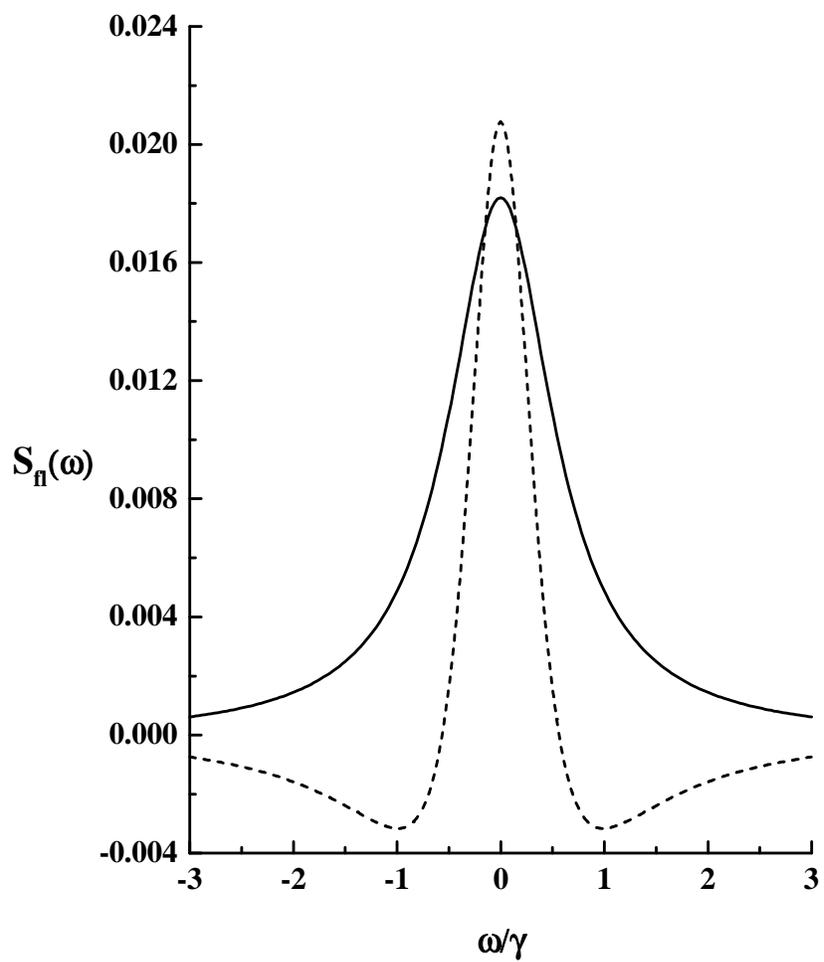
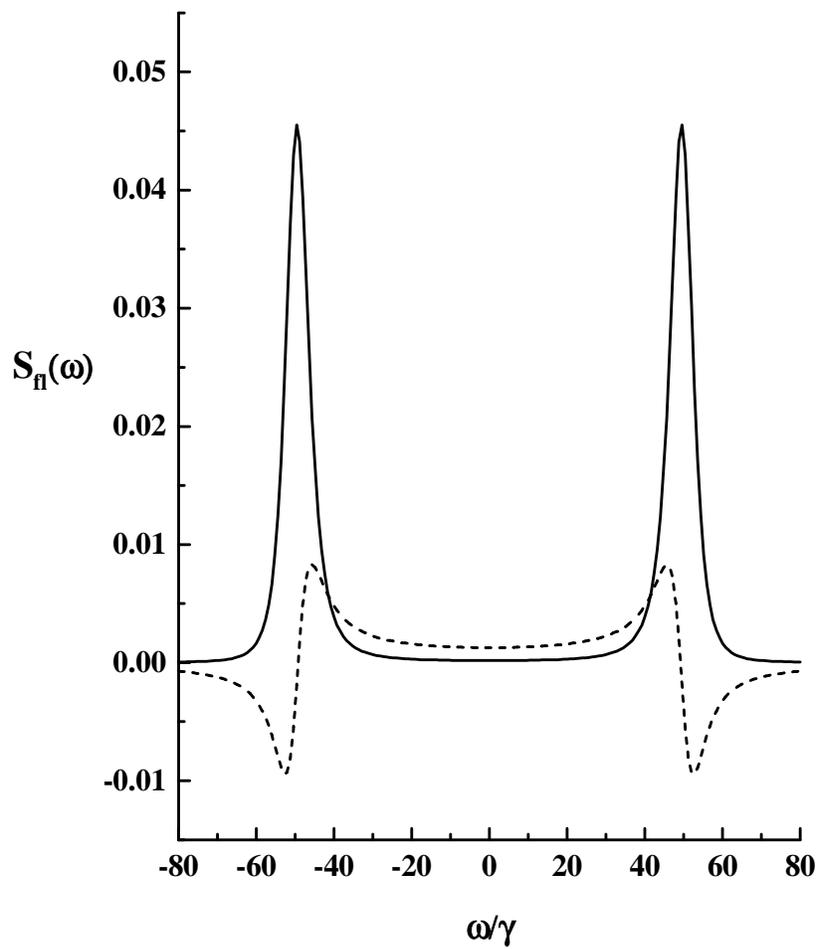
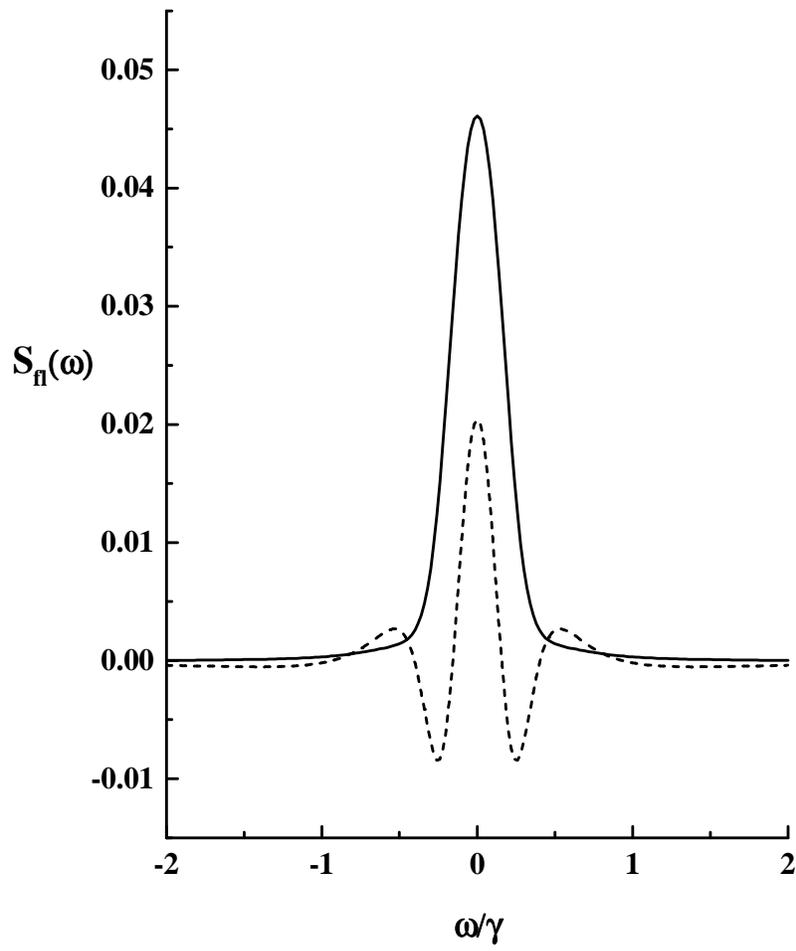
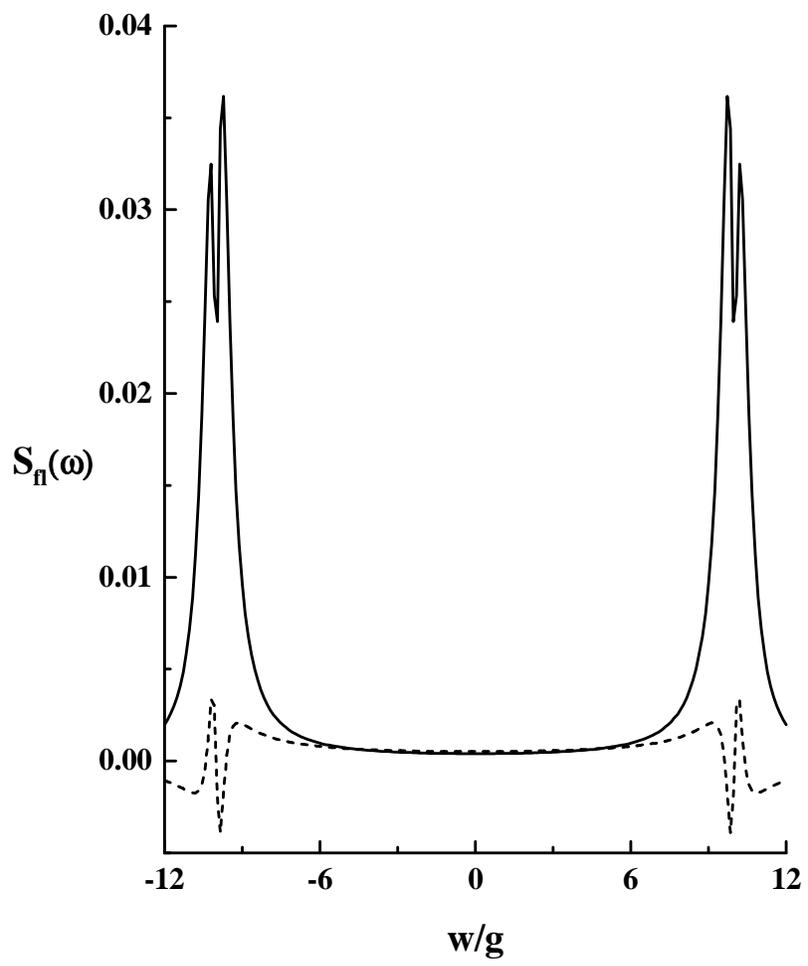
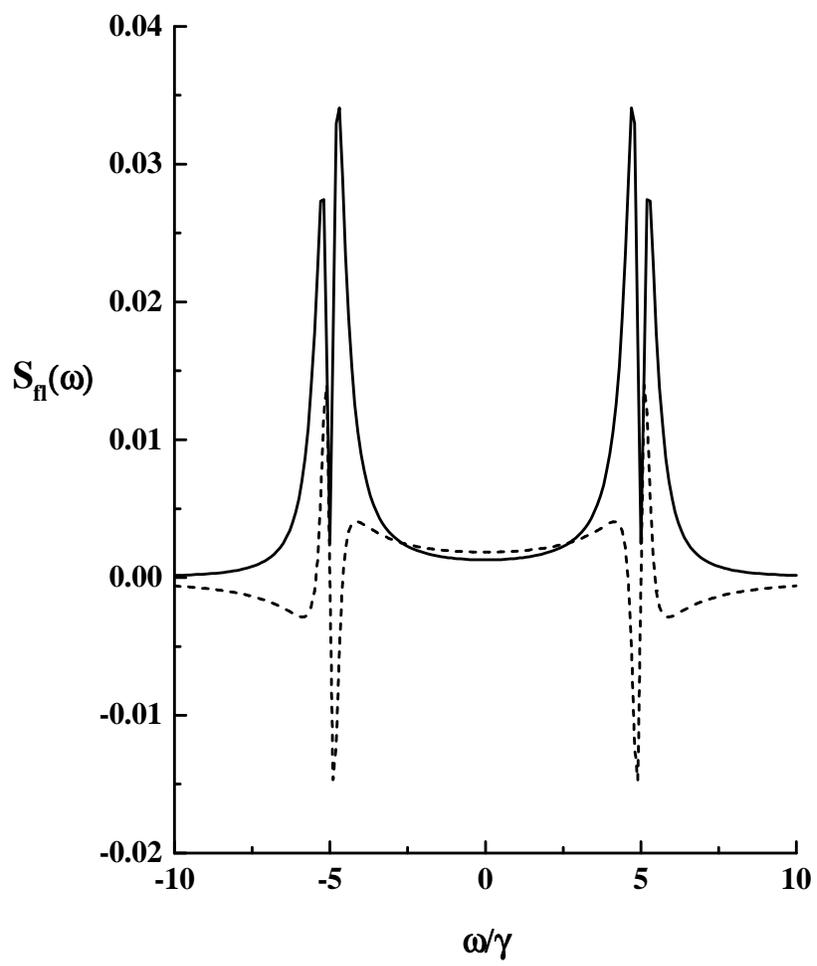
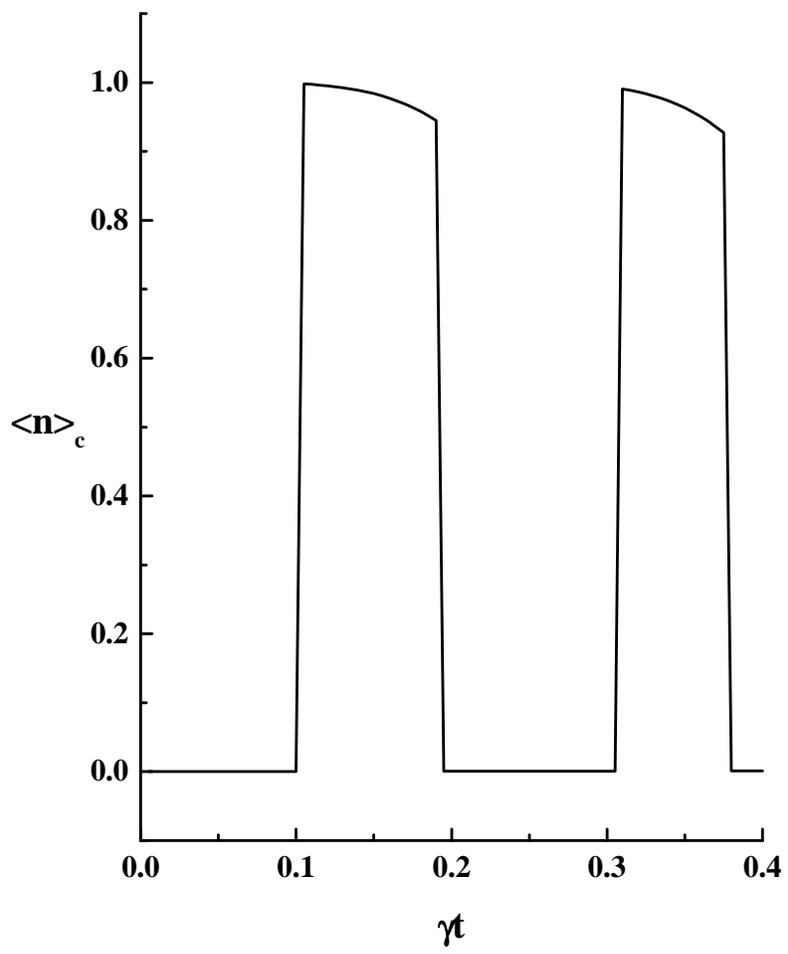
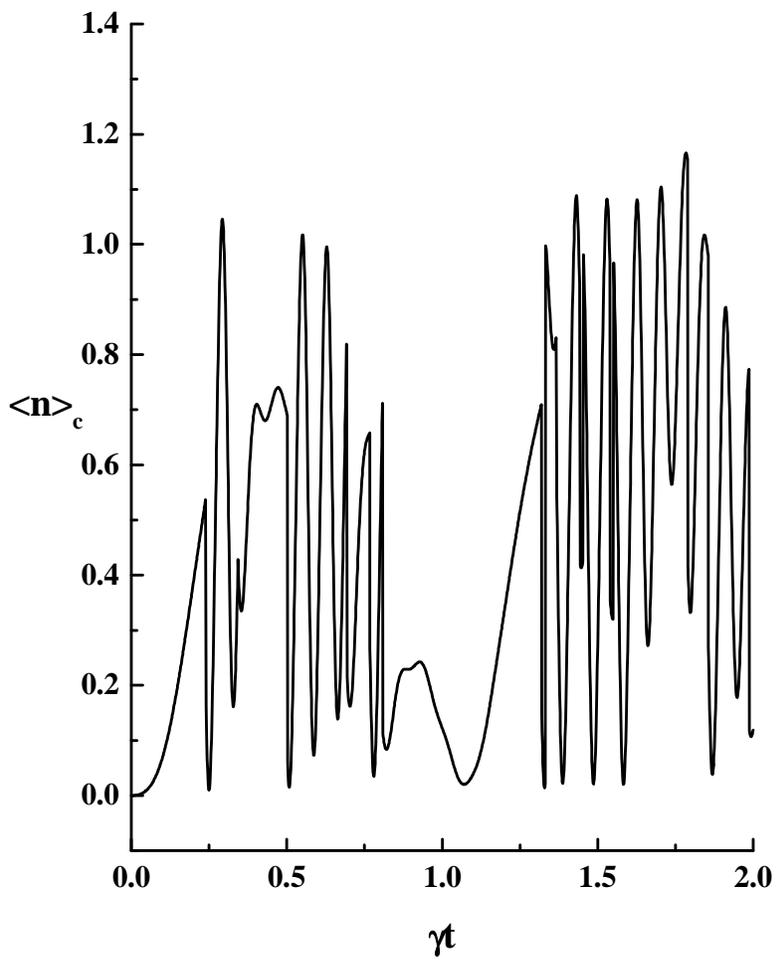
\end{document}